\documentclass[aps,pre]{revtex4-2}

\usepackage{amsfonts}
\usepackage{times}
\usepackage{amsmath}
\usepackage{graphicx}
\usepackage{amssymb}
\usepackage{xcolor,comment}

\begin{document}

\title{Dark solitons in the fractional NLS equation}

\author{Almudena P. M\'arquez}
\affiliation{Department of Mathematics, College of Engineering, University of Cadiz, 11519 Puerto Real, Cadiz, Spain}
\author{Jes\'us Cuevas-Maraver}
\affiliation{Grupo de F\'{\i}sica No Lineal, Departamento de F\'{\i}sica Aplicada I,
Universidad de Sevilla. Escuela Polit\'{e}cnica Superior, C/ Virgen de \'{A}frica, 7, 41011-Sevilla, Spain} 
\affiliation{Instituto de Matem\'{a}ticas de la Universidad de Sevilla (IMUS). Edificio
Celestino Mutis. Avda. Reina Mercedes s/n, 41012-Sevilla, Spain}
\author{Panayotis G.\ Kevrekidis}
\affiliation{Department of Mathematics and Statistics, University
of Massachusetts, Amherst,  01003-4515, MA USA} 
\affiliation{Department of Physics, University of Massachusetts Amherst, 01003, MA USA}
\affiliation{Department of Mechanical Engineering, Seoul National University, 1 Gwanak-ro, Gwanak-gu, Seoul 08826, South Korea}

\date{\today}

\begin{abstract}
In the present work we consider the subject of dark fractional 
solitary waves in the realm of generalized (fractional) forms of the
nonlinear Schr{\"o}dinger (NLS) equation. While earlier studies have examined such states in the realm of real field theories, we showcase 
the existence and stability of individual dark solitary waves in 
such NLS settings and subsequently turn to two-soliton solutions. 
We find different branches of such two-soliton solution equilibria
and contrary to the real field-theoretic setting {\it all} possible
branches of two-soliton equilibria are found to be potentially unstable,
although with different types of instabilities. Odd branches are potentially subject to oscillatory instabilities, while even branches are always  exponentially unstable. The dynamics that results from the instabilities is also examined and is found to potentially feature breathing characteristics. This prompts us to seek and find associated periodic (breathing) orbits that are also unprecedented in this context, to the best of our knowledge. The effective particle-like dynamics of the solitary waves also prompts us to seek ordinary differential equation (ODE) descriptions to the dark soliton interaction dynamics. These are shown to hold promise toward providing us with effective reduced order models, indicating some potential directions for further investigation.
\end{abstract}

\maketitle

\section{Introduction}

Recent experimental advances in nonlinear optics have opened new vistas
toward the exploration of dispersive wave models beyond the conventional second-order paradigm. In particular, within the last decade the ability to engineer dispersion landscapes has enabled controlled access to higher-order dispersive regimes~\cite{BlancoRedondoNC2016,RungeNP2020}. A landmark outcome of these efforts was the experimental observation of pure quartic solitons, which subsequently stimulated significant theoretical, numerical, and experimental work on their properties and extensions~\cite{deSterkeRungeHudsonBlancoRedondoAPL2021PQS,ParkerAcevesPhysD2021Multipulse,DeckerDemirkayaMantonKevrekidisJPA2020BiharmonicPhi4,TsoliasDeckerDemirkayaAlexanderParkerKevrekidisCNSNS2023NLSMixedDispersion,TamOL2019,TamPRA2020,BandaraPRA2021}. The exploration of dispersive dynamics has since expanded to include even higher-order regimes~\cite{QiangAlexanderdeSterkePRA2022SixthOrder,deSterkeBlancoRedondoOptCom2023EvenOrder,WidjajaQiangSkeltonNatCommun2025Universality}. Despite these developments, 
considerably less attention has been directed toward dark solitary-wave states~\cite{Alexander:22} (as well as domain wall ones~\cite{KumarLiMalomedPRE2022Dark}), even though such structures may possess interesting dynamical features.

These optical advances have more recently led to a broader conceptual framework in which dispersion is not restricted to discrete integer orders but can instead vary continuously. In this context, fractional dispersion provides a natural mathematical tool for interpolating between harmonic and biharmonic limits (but also, of course,
toward exploring beyond these). A key question then concerns how the structure and interactions of solitary waves are modified as the dispersion exponent changes smoothly between ---as well as outside of--- these regimes. Related issues have recently been explored in a different setting, namely for kink solutions in a real scalar field theory~\cite{Bob}.

The possibility of describing wave propagation using fractional dispersion is rooted in the broader realm of fractional calculus. Independently of the above-mentioned developments, and, in particular, in other areas of applications, fractional models have attracted growing interest, motivated by their ability to capture nonlocal effects in diverse settings. Examples include epidemiological dynamics~\cite{qureshi2020real}, anomalous diffusion processes in biological systems~\cite{IonescuCNSNS2017}, models for the spread of computer malware~\cite{singh2018fractional,azam2020numerical}, and nonlinear wave~\cite{cuevas} and economic dynamics~\cite{ming2019application}. Optical systems have been front and center in relevant developments in this direction as well~\cite{malomed}. The rapid growth of the field of fractional calculus has been documented in several influential monographs and review articles; see, for instance,~\cite{podlubny1999fractional,Samko,Mihalache2021}.

At the same time, substantial effort has been devoted to the mathematical analysis of fractional operators themselves~\cite{yavuz2020comparing,saad2018new}. Despite this progress, the theoretical framework is still evolving, partly because multiple definitions of fractional derivatives coexist in the literature~\cite{ortigueira2021two,ortigueira2021bilateral}. Among the various formulations, the Riesz fractional derivative is particularly appealing for physical applications~\cite{muslih2010riesz}. One reason is that it arises naturally as the continuum limit of lattice systems with long-range interactions~\cite{tarasov2006fractional}, thereby providing a transparent connection between microscopic lattice models and effective fractional continuum descriptions.

Although the mathematical literature on fractional operators is extensive, experimental realizations of fractional-order wave dynamics have historically been scarce. Early steps toward such realizations appeared in optics, where the fractional Schr\"odinger equation was proposed as a model for optical propagation~\cite{Longhi:15} and later implemented experimentally in the linear regime~\cite{malomed}. More recently, direct experimental demonstrations of fractional dispersion have been achieved in optical platforms. These include temporal-domain implementations of the fractional Schr\"odinger equation~\cite{LiuNatCommun2023FSE} as well as nonlinear propagation governed by fractional derivatives, as reported in the pioneering work of~\cite{HoangNatCommun2025FractionalDerivative}. A further experimental development involves tunable dispersion relations in Fourier space, allowing for interpolation between the $\alpha=1$ Hilbert-type nonlinear Schr\"odinger equation and the corresponding Nyquist-type variant, although without explicitly varying the fractional exponent $\alpha$ considered here~\cite{martijn25}.

In the present work we are considering the far less studied case of dark
solitary waves in fractional NLS models. 
{For the latter, it is relevant to note that the recent work
of~\cite{Kusdiantara2026} examined the realm of both bright and dark solitary
waves in a harmonic trap. This is to be contrasted with the absence of such
confinement in the present setting, i.e., the exploration of such solitary
waves (and of molecules thereof) in free space.}
Upon brief consideration
of the single solitary wave setting, we turn our attention to two-soliton
molecules and study their stability. We find that multiple such states
---in analogy with the real field-theoretic problem---. However, we 
showcase (and explain) their fundamentally different stability characteristics,
which imply that every such configuration is potentially unstable. The
odd ones among them may feature oscillatory instabilities, while the even
ones bear a real pair of eigenvalues. The dynamics of these instabilities
are accordingly considered and it is found that in the case of the oscillatory
instability (Hamiltonian Hopf bifurcation) a breathing mode arises.
This, in turn, prompts us to identify such breathing waveforms
as time-periodic solutions of the system and to perform their stability analysis.
Finally, in light of the effective particle nature of the dark solitonic
waveforms, we seek to characterize their dynamics both in the single soliton
and in the molecule setting by means of a variational (reduced order)
characterization. 

Our presentation is structured as follows.
In section II we present the mathematical backdrop of the relevant
model introducing also the fractional (Riesz) derivative of interest.
In section III we proceed to analyze the model numerically both at the
level of a single and at that of multiple (steady or time periodic)
dark solitons. In section IV we seek to leverage a variational
approach towards describing the effective solitonic dynamics,
while in section V we summarize our findings and present our
conclusions, as well as a number of directions for future study.

\section{The model}

We consider a defocusing NLS equation with a dispersive term in the generalized 
form of Riesz derivative~\cite{malomed2}:
\begin{equation}\label{eq:dyn}
    i\partial_t\psi=-\partial^\alpha_x\psi+|\psi|^2\psi
\end{equation}
with
\begin{equation}
    \partial^\alpha_x\psi(x,t)=-\frac{1}{2\pi}\int_{-\infty}^\infty \mathrm{d}k |k|^\alpha \int_{-\infty}^\infty \mathrm{d}\xi \mathrm{e}^{ik(x-\xi)} \psi(\xi,t)
\end{equation}
being the Riesz fractional derivative. This is a disguised spectral derivative in Fourier space, which ---based on the above form of its realization--- 
enforces periodic boundary conditions in the domain. When the fractionality or L\'evy index $\alpha$ is equal to 2, the classical Laplacian is recovered, while
for instance the case of $\alpha=4$ represents the biharmonic limit
and continuations between the two become possible via the variation
of $\alpha$~\cite{Bob}.

As is customary in NLS settings, we look for stationary solutions $\psi(x,t)=\phi(x)\exp(i\omega t)$ fulfilling:
\begin{equation}\label{eq:stat}
    -\omega\phi+\partial^\alpha_x\phi-\phi^3=0.
\end{equation}

Once the relevant standing waves are identified, spectral stability can be studied by introducing a perturbation $\xi(x,t)$ to the wavefunction $\psi(x,t)$ in Eq. (\ref{eq:dyn}),
using the decomposition:
\begin{equation}
\psi(x,t)=e^{i\omega t}\left[\psi_0(x,t)+\xi(x,t)\right].
\end{equation} 
Considering the leading order dynamics of $\xi(x,t)$, one can reach to the following
linearization equation:
\begin{equation}\label{eq:var}
i\partial_t\xi=-\partial^\alpha_x\xi+\omega\xi+2|\psi_0|^2\xi+\psi_0^2\xi^*.
\end{equation}

In what follows, we will assume that the solutions $\phi_0$ of Eq.~(\ref{eq:stat})
are real.
Then, for stationary solitons~\footnote{For simplicity, we will use the terminology
of ``solitons'' here ---as is common in the Mathematical Physics community---, 
although this is with the understanding that these will be true solitons mathematically
only in the limit of $\alpha=2$.}, $\psi_0(x,t)=\phi(x)$ and, by expressing the perturbation as $\xi(x,t)=a(x) e^{\lambda t}+b^*(x) e^{\lambda^* t}$, one can write a spectral problem in the form
\begin{equation}
\lambda\left(\begin{array}{c}a(x)\\b(x)\end{array}\right)=-i
\left(\begin{array}{cc}L_1 & L_2 \\-L_2 & -L_1 \end{array}\right)
\left(\begin{array}{c}a(x)\\b(x)\end{array}\right).
\end{equation} 
The block operators $L_i$, $i=1,2$ are of the form
\begin{equation}
\begin{split}
L_1 &= \omega-\partial^\alpha_x+2|\phi|^2,  \\
L_2 &= \phi^2.
\end{split}
\end{equation} 

Equation (\ref{eq:dyn}) also supports the existence of breathers, which resemble oscillating bound dark soliton states. In order to seek such breathers, the following Fourier series expansion
\begin{equation}\label{eq:Fourier}
\psi(x,t)=e^{i\omega t}\sum_{k=-K}^K z_k(x)e^{ik\Omega t}
\end{equation}
can be introduced in Eq. (\ref{eq:dyn}) and the following ODE set is attained
\begin{equation}\label{eq:stat_Fourier}
    \mathcal{G}_k\equiv-(k\Omega+\omega)z_k+\partial^\alpha_x z_k+\sum_\ell\sum_{\ell^\prime}z_kz^*_\ell z_{k-\ell+\ell^\prime}=0.
\end{equation}

Once a solution of this set of ordinary differential equations (in space) is obtained, 
the stability of the resulting breathers is determined by means of Floquet analysis. To this aim, the linearization equation (\ref{eq:var}) is used for solving the tangent map
\begin{equation}
\left[\begin{array}{c}
  \textrm{Re}(\xi(x,T)) \\[2mm] \textrm{Im}(\xi(x,T)) \\
  \end{array}\right]
  =\mathcal{M}
  \left[\begin{array}{c}
  \textrm{Re}(\xi(x,0)) \\[2mm] \textrm{Im}(\xi(x,0)) \\
  \end{array}\right]
\end{equation}
with $T=2\pi/\Omega$. As the system is Hamiltonian, breathers are stable whenever the spectrum of the monodromy matrix (whose eigenvalues are also known as Floquet multipliers) $\{\Lambda\}$, lie on the unit circle. In what follows, we will explore
the stability of both solitary waves and their pairs, as well as that of breathers.

\section{Numerical results}

\subsection{L\'evy index $\alpha<2$. Dark solitons}

\begin{figure}
\begin{center}
\begin{tabular}{cc}
\includegraphics[width=0.45\textwidth]{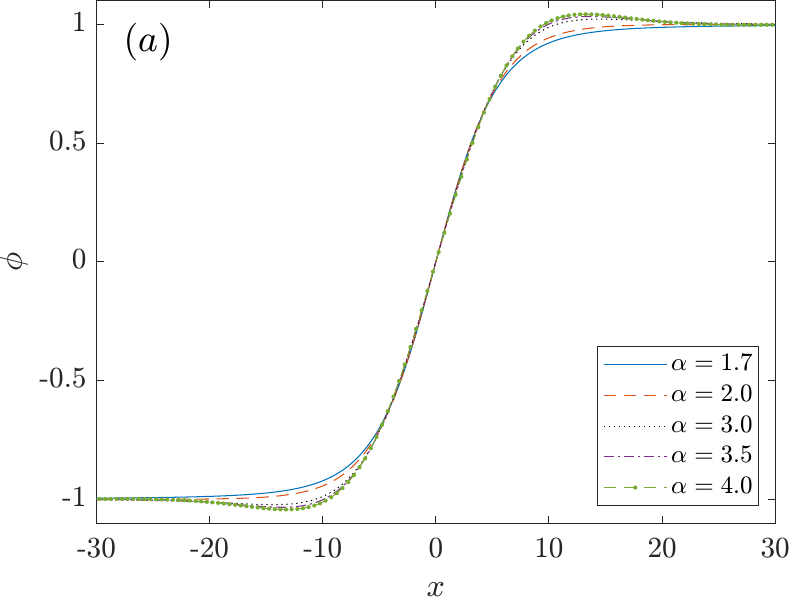} &
\includegraphics[width=0.45\textwidth]{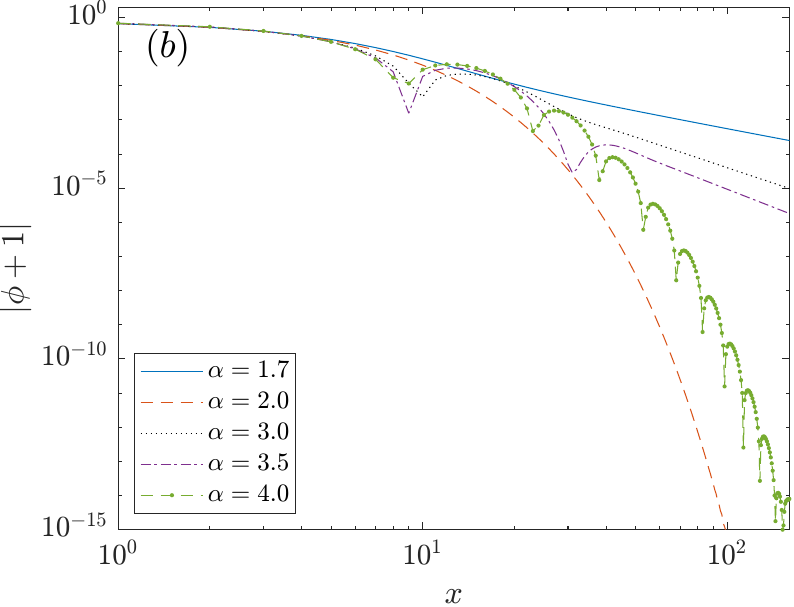}
\end{tabular}
\end{center}
\caption{Profile of dark solitons in linear (a) and logarithmic scale (b) for different values of $\alpha$. Blue (solid), red (dashed), black (dotted), purple (dash-dotted), and green (thick dash-dotted) lines correspond, respectively, to $\alpha=1.7$, $\alpha=2$, $\alpha=3$, $\alpha=3.5$, and $\alpha=4$.}
\label{fig:profiledark}
\end{figure}

We start by considering the stability and dynamics of dark solitons when $\alpha<2$. Such dark soliton solutions exist when $\alpha<2$ and $\omega<0$ (without loss of generality, we take $\omega=-1$). Due to the periodicity of the boundary conditions imposed by the Fourier spectral derivative, one must consider a dark soliton pair so that the members of the pair are far away. In the Laplacian limit $\alpha=2$, 
a single stationary dark soliton has a profile of a hyperbolic tangent, i.e.,
$\psi_0(x)=\tanh(x/\sqrt{2})$.
When seeking to compute such states numerically, the periodic nature of the 
boundary conditions does not allow for the realization of a single one, hence
we utilize an ansatz of a dark solitonic pair in the form:
\begin{equation}\label{eq:twosolitons}
\phi(x)=\tanh\left(\frac{x+\delta}{\sqrt{2}}\right)-\tanh\left(\frac{x-\delta}{\sqrt{2}}\right)-1
\end{equation}
where $2\delta$ is the separation between solitons ($\delta$ is also denoted as the soliton's offset). In our particular case, we have taken $\delta={\rm O}(100)$ so that the solitons are sufficiently well separated. We observe in 
Fig.~\ref{fig:profiledark} that as one departs from the exponential tail setting of 
$\alpha=2$, the solitary waves acquire a monotonic tail in the case
of $\alpha<2$, while they become non-monotonic for any $\alpha>2$. 
This is in line with the observations of~\cite{AtanasPanos} since the
steady state problem of Eq.~(\ref{eq:stat}) is shared between the
real field-theoretic $\phi^4$ problem of~\cite{AtanasPanos} and the
NLS considerations herein. As $\alpha \rightarrow 4$, the oscillations
in the tail increase, tending to an infinite limit as $\alpha$ approaches
the biharmonic limit. 

The single dark solitary wave of the fractional NLS problem is found
to be generically stable for the values of $\alpha \in [1.5,4]$ that
we have considered herein. 
The spectrum of the solitary wave always bears two pairs of
eigenvalues at the origin, due to the translational and phase 
symmetries of the standard NLS (that are preserved here), while
the continuous spectrum encompasses the entire imaginary axis~\cite{Carretero24}.
However, it is important to add here a note of caution for numerical 
practitioners. The fact that the computation of the soliton requires
(given the periodic boundary conditions) the inclusion of a second 
structure, per the ansatz of Eq.~(\ref{eq:twosolitons}) produces some
spurious effects. Practically, each of the solitons carries a phase
and a translational mode pair, hence any configuration with multiple
solitons will bring about multiple such pairs. While in the Laplacian
limit of the regular NLS and exponential solitonic tails these waves will
be blissfully unaware of each other's presence (for the numerical distances
considered) and the relevant spectra will represent a ``doubling'' of the
spectrum of a single soliton, as $\alpha$ is varied and the long-range
power-law tails come into effect, this is no longer the case. Then, 
while the (global) translational mode of the pair, representing the 
in-phase translation of the two-solitons is still a global symmetry,
the out-of-phase motion between the waves will no longer be reflected
in a zero eigenvalue. Practically, this will lead to {\it spurious}
instability eigenmodes as this eigenvalue ``penetrates'' the continuous
spectrum of the imaginary axis and collides with modes thereof producing
an eigenvalue quartet (similar to the one observed, e.g., in~\cite{johkiv}). Our numerical stability results 
reflected this spurious instability, notably for values of 
$\alpha<1.7$. 

\subsection{L\'evy index $\alpha>2$. Bound states and breathers}

\begin{figure}
\centering
    \includegraphics[width=0.45\textwidth]{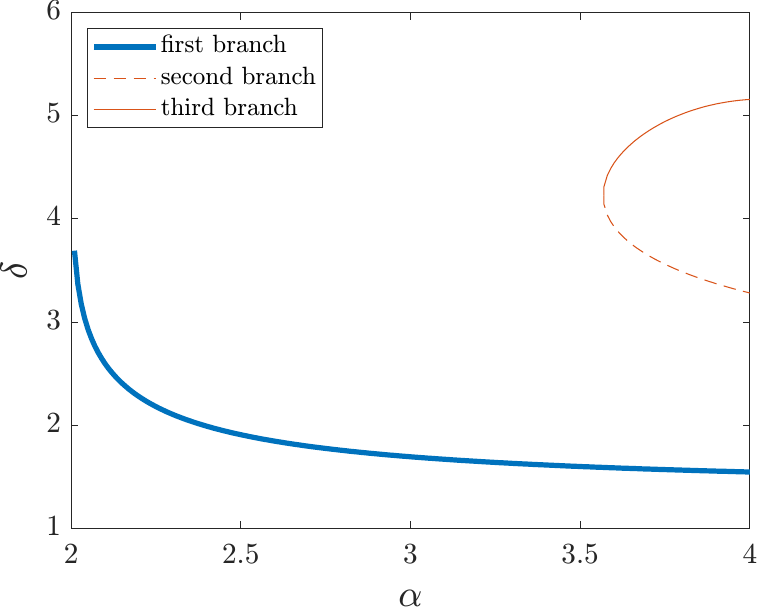}
    \caption{Dependence of the equilibrium distance between the dark soliton centers in a bounded pair $\delta$ on the L\'evy index $\alpha>2$. Notice the existence of two families (or three branches) interpolating between $\alpha=2^+$ and $\alpha=4$. 
    Additional branches emerge in the immediate vicinity of the biharmonic limit
    of $\alpha=4$, although given their exponential proximity to the biharmonic 
    limit, we will not explore them here. Branches in continuous (dashed) lines correspond to oscillatorily (exponentially) unstable solutions.}
    \label{fig:bifbound}
\end{figure}

\begin{figure}
\begin{center}
\begin{tabular}{cc}
\includegraphics[width=0.45\textwidth]{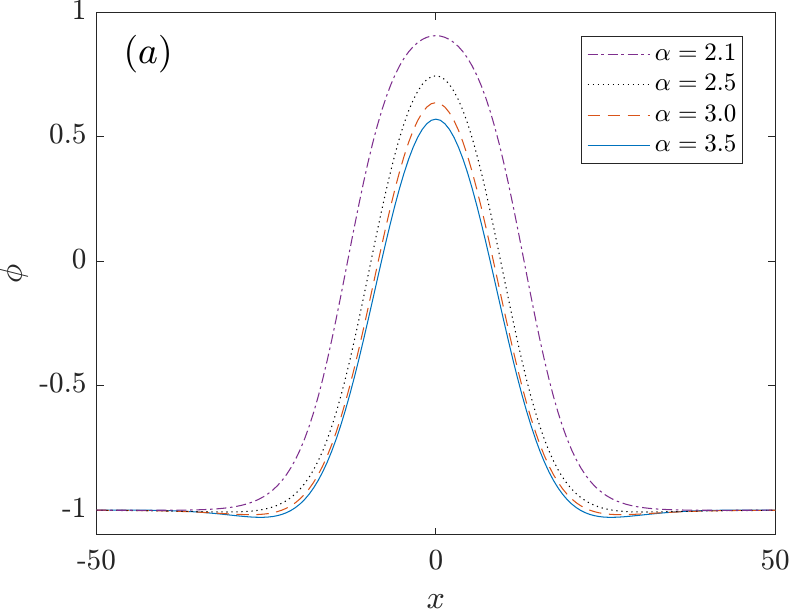} &
\includegraphics[width=0.45\textwidth]{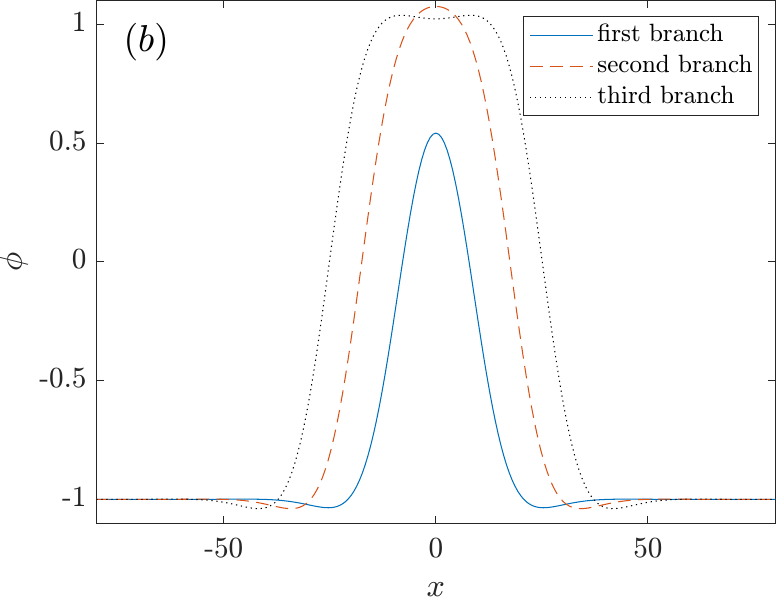} \\
\includegraphics[width=0.45\textwidth]{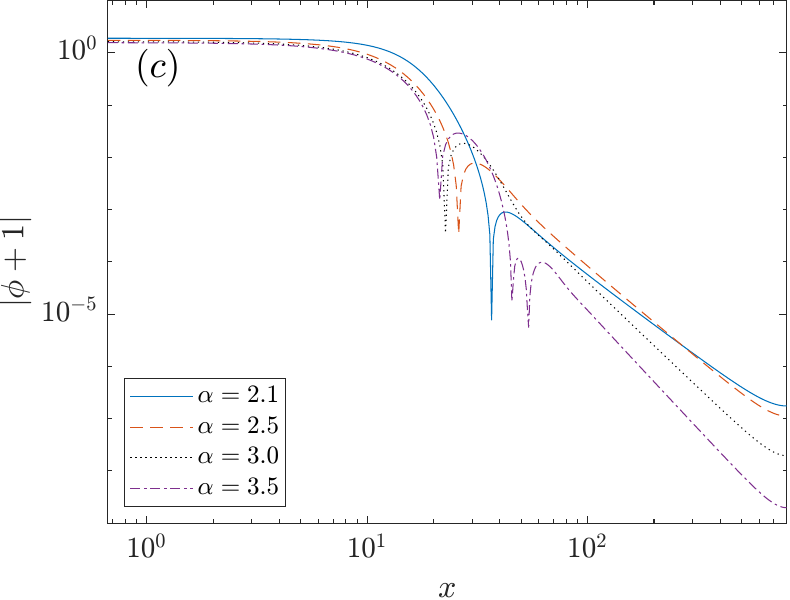} &
\includegraphics[width=0.45\textwidth]{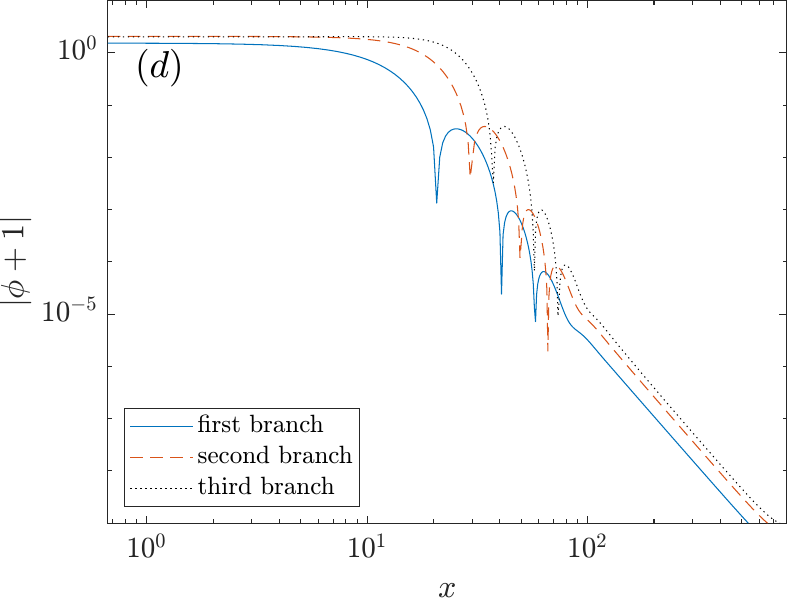} \\
\end{tabular}
\end{center}
\caption{(a) Profiles of the bound states in the first branch for $\alpha=3.5$ (blue solid line), $\alpha=3$ (red dashed line), $\alpha=2.5$ (black dotted line), and $\alpha=2.1$ (purple dash-dotted line). (b) Profiles of the bound states for the lowest (blue solid line), second (red dashed line) and third (black dotted line) branches at $\alpha=3.8$. (c)-(d) Profiles in logarithmic scales corresponding to (a) and (b), respectively.}
\label{fig:profilesbound}
\end{figure}

When $\alpha>2$, as indicated above (as well as in~\cite{AtanasPanos}),
the tail of the dark solitons acquires a zero-crossing.
Accordingly, the force between the two dark solitons which is
dictated by the relevant tail~\cite{Manton1979EffectiveLagrangian,Manton_2019},
also acquires a zero crossing. This implies that for distances shorter
than this vanishing point, the force remains repulsive (as it is
for standard NLS dark solitons~\cite{KIVSHAR1995353}), whereas for distances
longer than that, the force is now attractive. Accordingly, dynamically, this
is expected to be a stable equilibrium point, based on the corresponding force field.
In fact, as $\alpha$ is further increased, there exist several families of bound states in which the separation (or offset) between the solitons is a function of $\alpha$, as one can see in Fig.~\ref{fig:bifbound}; see also~\cite{Bob}. We were able to find three different branches interpolating between $\alpha=2^+$ and $\alpha=4$, although 
progressively more should arise, en route to the infinitely many such bound
states in biharmonic limit of $\alpha=4$. For computing those bound states and their stability properties, we have used a domain $[-L,L]$ with $L=160$ and a discretization parameter $h=2/15$. Figure~\ref{fig:profilesbound} shows the profile of the bound state of the lowest branch for selected values of $\alpha$ and also the profiles of $\alpha$ for the three branches at the same value.

\begin{figure}
\centering
\begin{tabular}{cc}
\includegraphics[width=0.45\textwidth]{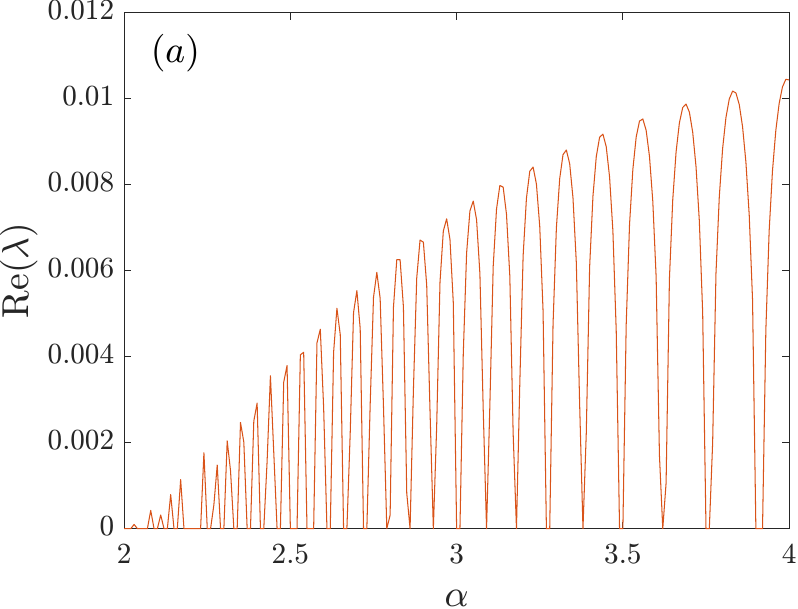} &
\includegraphics[width=0.45\textwidth]{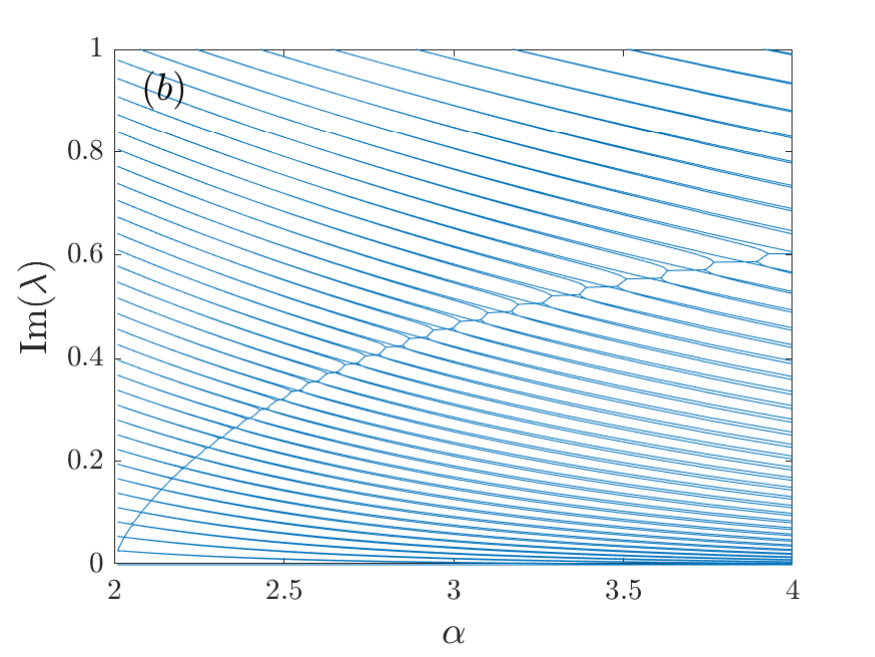} \\    
\end{tabular}
    \caption{Dependence with $\alpha$  of the real (a) and imaginary (b) part of the spectrum for the lowest branch of bound states for $\alpha>2$.}
    \label{fig:stab1stbranch}
\end{figure}

The three families, for which the $\delta-\alpha$ curve coincides with that for the fractional $\phi^4$ mode shown in~\cite{Bob}, manifest opposite stability properties ---to the latter model--- in what concerns exponential instabilities. That is, the lower branch of the second family (i.e., the second branch) is the only one presenting solutions which are exponentially unstable. The solutions belonging to the other branches are free from exponential instabilities but, on the contrary, present oscillatory instabilities, as we explain in what follows. The lowest family, consisting of an isolated branch of solutions, is the only one that exists in the neighbourhood of $\alpha=2$. 
The spectral picture of the linearization associated with this
family reveals three pairs of localized eigenmodes existing, namely two pairs of 
neutral modes at $\lambda=0$ (corresponding to symmetries associated with the dark solitons) 
and a third pair responsible for the interaction between the solitons (dubbed as {\em interaction mode}), whose associated eigenvalue $\lambda_i$ is imaginary. 
When $\alpha\rightarrow2^+$, $\lambda_i\rightarrow0$, leading to a bifurcation precluding from the existence of soliton molecules for $\alpha\leq2$. When $\alpha$ is increased, the imaginary part $|\lambda_i|$ of the ``interaction eigenvalue'' 
also increases and collides with the linear extended (continuous spectrum) modes (see Fig.~\ref{fig:stab1stbranch}) through Hopf bifucations, leading to an eigenvalue quartet and the solution becomes oscillatorily unstable. Subsequent increase in $\alpha$ leads to the disappearance of the quartet via an inverse Hopf bifurcation 
and the solution becomes stable. This mechanism is repeated along the entire the solution family when $\alpha$ is varied. In other words, the instabilities manifest as bubbles, again in analogy with what is presented in~\cite{johkiv}. The cause of such bubbles are the gaps in the continuous spectrum because of the finite domain size. Consequently, we anticipate that in an infinite domain, the bound state would be unstable for every $\alpha>2$, analogously to what
is conjectured for the infinite lattice size limit in~\cite{johkiv}.

\begin{figure}
\centering
\begin{tabular}{cc}
    \includegraphics[width=0.45\textwidth]{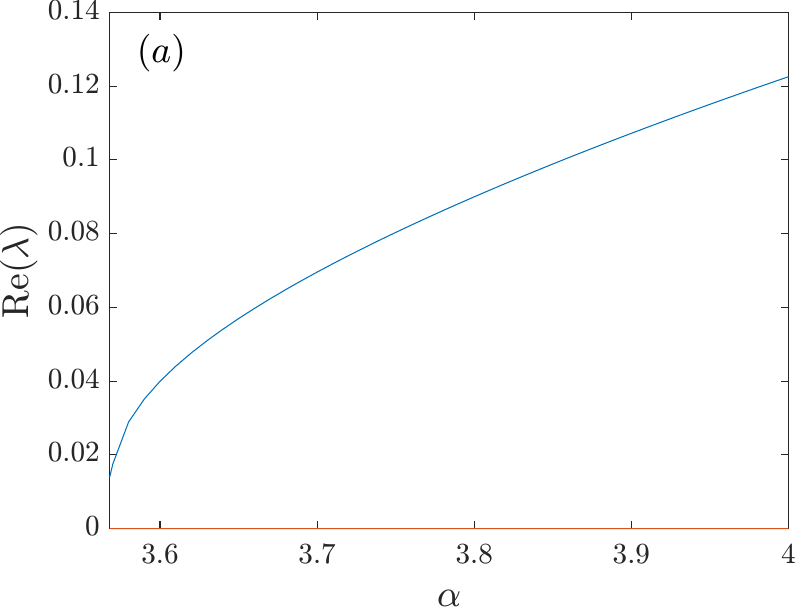} &
    \includegraphics[width=0.45\textwidth]{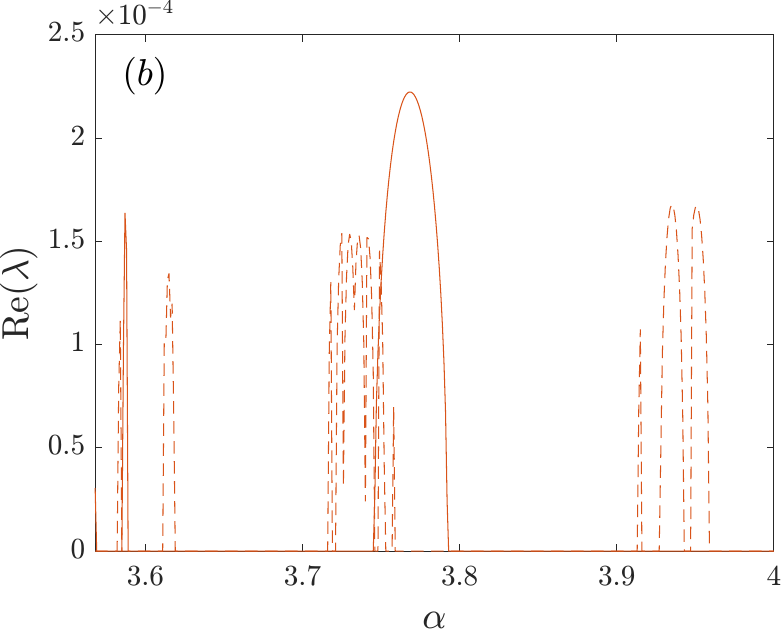}  \\    
\end{tabular}
    \caption{Dependence on $\alpha$ of the real part of the spectrum for the lower (a) and upper (b) branch of the second family of bound states. Dashed lines in the right panel correspond to a larger domain than full lines; in particular, the domain is $[-L,L]$ with $L=320$ and a discretization parameter $h=2/15$.}
    \label{fig:stab2ndfamily}
\end{figure}

Turning now to the second family, we find that the lower branch is exponentially unstable for every $\alpha$, as the linearization
picture reveals an interaction eigenmode pair that is real. In this case, when $\alpha$ approaches the bifurcation point, $\lambda_i\rightarrow0$; on the contrary, the upper branch's spectrum is similar to that of the lowest family so that the upper branch features oscillatory instability bubbles (alternating with windows
of stability). Figure~\ref{fig:stab2ndfamily} shows the real part of the eigenvalues of the bound states for this family; in the upper branch we have considered two different domain lengths, so that one can see how the number of bubbles increases when the domain increases. 
As the domain size changes, the 
``quantization'' of the continuous spectrum modes (with which the 
mode pertaining to the out-of-phase motion of the two dark solitons 
collides to produce a resonance), it is expected to affect, accordingly,
the precise location of the unstable windows, in a way similar to what is
observed in~\cite{johkiv} (for different domain sizes; see Fig.~2 therein).

It is instructive to consider the nature of interaction between
the constituent dark solitons for all of these families.
In the case of a transition of this interaction force 
from an attractive to a repulsive one
(as the displacement is increased), clearly we are dealing with a saddle
point and, hence, the dynamics is naturally (exponentially) unstable
as in the example of the lower branch of the second family (i.e., the second
branch).
The transition from repulsive to attractive force, on the other hand,
in the first and third branch
should amount to a restoring force leading, in principle, to oscillations and
an associated oscillation frequency. However,
the resonance of the frequency of these oscillations with modes
of the continuous spectrum (which encompasses the entire imaginary 
axis~\cite{johkiv,PelinovskyKevrekidis2008}) induces a path for the transfer of the
energy from this internal vibration mode to extended background modes.
As such, this resonance phenomenon leads to a complex eigenvalue
quartet and to the oscillatory, anti-damped evolution of the dark
soliton position, in line with the works of~\cite{johkiv,PelinovskyKevrekidis2008}.

It is worthwhile to connect here this stability picture with the expectation of
the force field between the two dark solitons. As indicated above, when
a single zero crossing of the force exists, the inter-soliton 
force  alternates from repulsive
(for shorter distances) to attractive (for longer ones) associated with 
oscillatory motion around this equilibrium. When another (second) equilibrium
emerges then around that, the force will alternate from attractive to repulsive,
suggesting a dynamically unstable fixed point. Then, for a subsequent fixed
point again the transition from repulsive to attractive forces will suggest
stability and so on. This relates odd fixed points to stable (oscillatory)
dynamics and even ones to unstable (saddle-point) ones. For the latter,
the connection to the real eigenvalue pairs and exponential instability
is immediate. For the former ones, the oscillatory dynamics implies
an imaginary eigenvalue. This, in turn, collides with the continuous
spectrum ones in a resonance amounting to a Hamiltonian Hopf 
bifurcation in a way similar to what is discussed in~\cite{PelinovskyKevrekidis2008},
and produces the oscillatorily unstable dynamics showcased in Fig.~\ref{fig:stab1stbranch}, as well as in
Fig.~\ref{fig:stab2ndfamily}(b).

\begin{figure}
\centering
\begin{tabular}{cc}
    \includegraphics[width=0.45\textwidth]{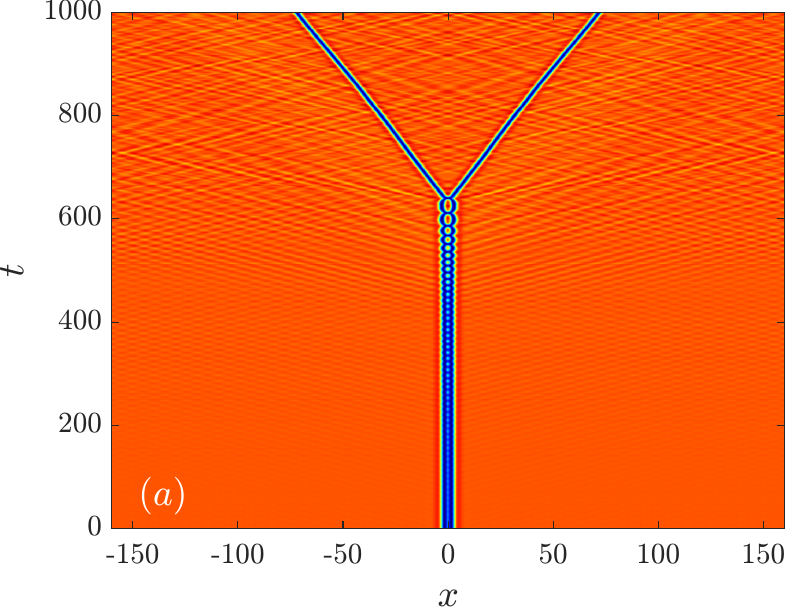} &
    \includegraphics[width=0.45\textwidth]{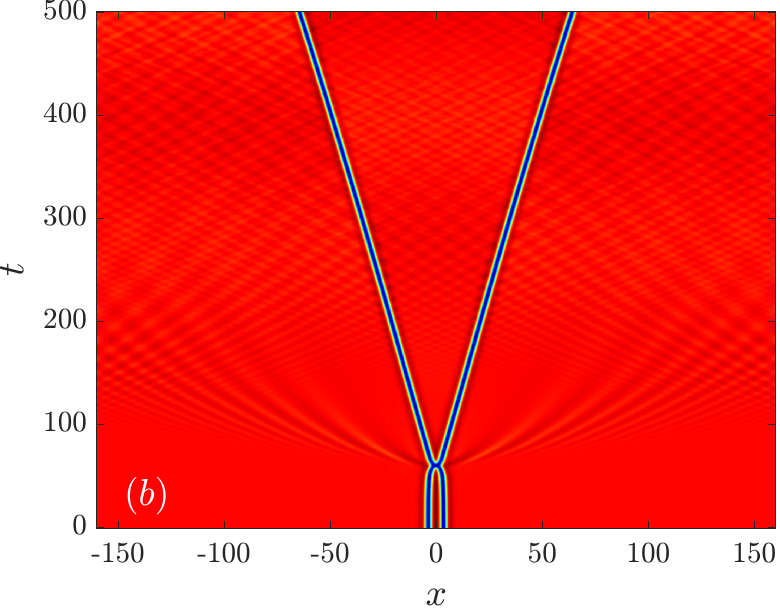}\\    
\end{tabular}
\caption{{Time dependence of the density for a bound state in the first branch with $\alpha=3.8$ (a), and in the lower portion of the second branch with $\alpha=3.7$ (b).}}
\label{fig:simbound}
\end{figure}

We have monitored the dynamical instabilities of the waveforms associated
with the two branches of ``solitonic molecules''; see Fig.~\ref{fig:simbound}. For the first branch, corresponding to oscillatory instabilities, a breathing bound state is formed (as is expected by the oscillatory nature of the instability and the
associated complex eigenvalues) before, eventually, the dark solitons indefinitely separate;
cf. Fig.~\ref{fig:simbound}(a).
In the second case, corresponding to exponential instabilities and real
eigenvalues (for the lower portion of the second branch in Fig.~\ref{fig:bifbound}), 
depending on the perturbation, if the inter-soliton distance is initially
made larger than the fixed point distance, they will immediately repel
and indefinitely separate. If, as in the middle panel, the distance between
them is initially decreased, they will attract (in the case of
Fig.~\ref{fig:simbound}(b)) until they collide and
then they will go their separate ways. All of these dynamics are
consistent with the effective particle interaction picture put forth
above. {For the upper portion of the second branch, the growth rate is very small. As a consequence, the time needed for the emergence of an appreciable instability 
manifestation is very large, rendering the computation time practically prohibitive, 
and hence we do not show associated results herein.}

\begin{figure}
\centering
\begin{tabular}{cc}
    \includegraphics[width=0.45\textwidth]{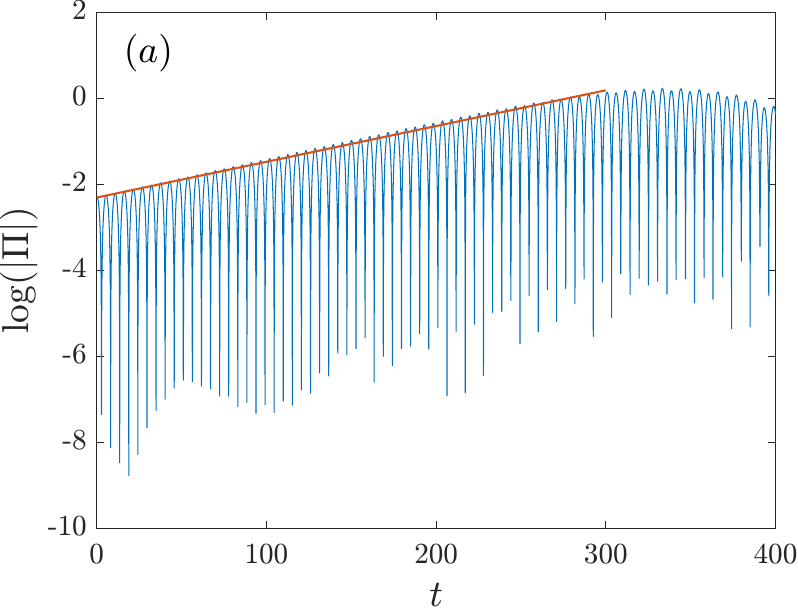} &
    \includegraphics[width=0.45\textwidth]{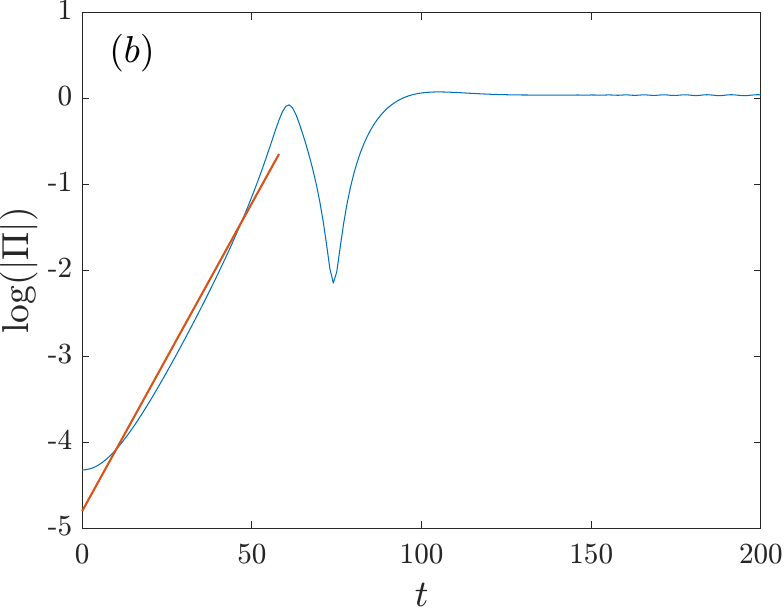} \\
    \includegraphics[width=0.45\textwidth]{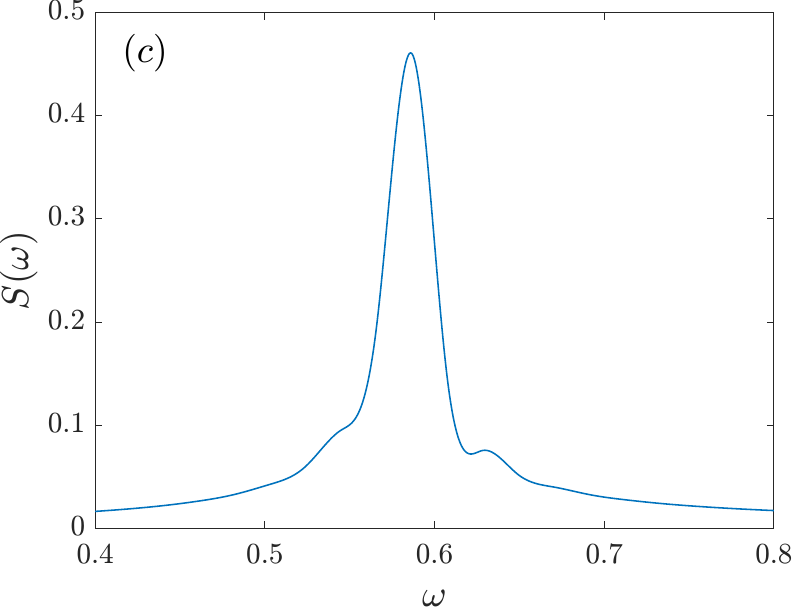} & \\
\end{tabular}
\caption{{The blue solid line in the plots of (a) and (b) corresponds to the evolution of the logarithm of the projection defined in Eq. (\ref{eq:projection}) for the bound states whose dynamics is displayed in Fig.~\ref{fig:simbound}. The red line is the fitting of the curve to $\log|\Pi|\sim\lambda' t$. The plot of (c) shows the Fourier spectrum of $\mathrm{Re}(\Pi)$. In (a)
(for $\alpha=3.8$), the unstable eigenvalue was $\lambda=0.0088\pm0.5868\mathrm{i}$ and the numerical growth rate and the oscillation frequency were found to be $\sim0.0082$ and $\approx0.58$, respectively.
For (b), associated with an exponential instability of the lower portion
of the second branch for $\alpha=3.7$,
the corresponding values were $\lambda=0.0696$ and $\lambda'\approx0.0715$. The domain is $[-L,L]$ with $L=160$ and a discretization parameter $h=2/15$.}}
\label{fig:projsimbound}
\end{figure}

In order to check that the dynamics stems from the non-zero real part eigenvalue of the spectrum, we calculate the {projection of the solution $\psi(x,t)$ minus the unperturbed solution $\tilde{\psi}(x,t)$} onto the corresponding eigenvector.
This is defined as
\begin{equation}\label{eq:projection}
    \Pi(t)=\int\left[\psi(x,t)-\tilde{\psi}(x,t)\right]\left[a^*(x)+b(x)\right]\mathrm{d}x.
\end{equation}
{We accordingly verify that, for short times, the projection grows as $|\Pi(t)|\sim\mathrm{e}^{\lambda' t}$, with $\lambda'$ being the growth rate associated to the instability. Figure~\ref{fig:projsimbound} shows the evolution of $\log|\Pi(t)|$ for the first stages of the instabilities of the bound states shown 
in Fig.~\ref{fig:simbound}.
In the case of the top left panels, there exists an oscillatory growth and
its linear portion in the semi-logarithmic graph captures the real part of the relevant eigenvalue
while its oscillatory character
is reflected in the imaginary part of the relevant growth. {The frequency of such oscillations is observed in the Fourier spectrum of the bottom left panel.}
{Here, too, we find good agreement between the linear stability prediction
and the PDE's dynamical manifestation.}
In the right panel, a genuine exponential growth, reflecting the real eigenvalue
of the associated state (in the lower portion of the 2nd branch), is observed.}

\begin{figure}
\centering
\begin{tabular}{ccc}
    $\Omega=0.015$ & $\Omega=0.02$ & $\Omega=0.025$ \\
    \includegraphics[width=0.3\textwidth]{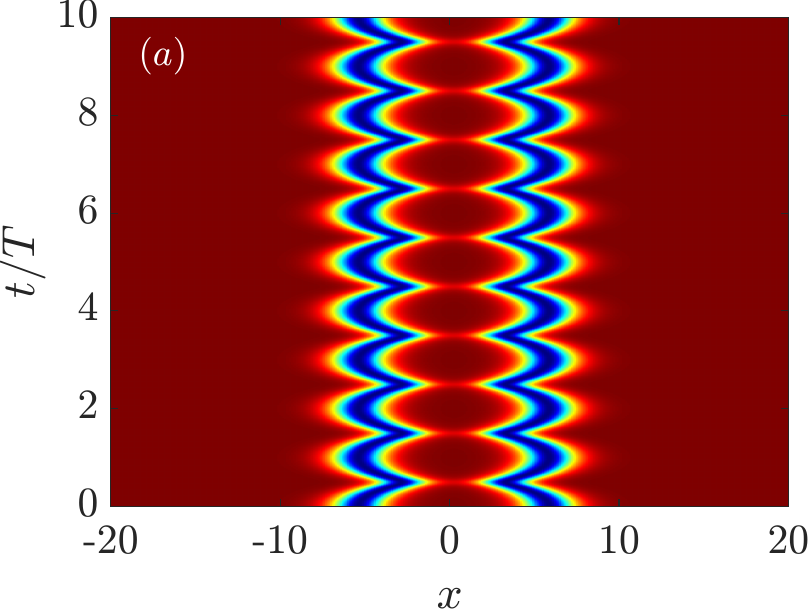} &
    \includegraphics[width=0.3\textwidth]{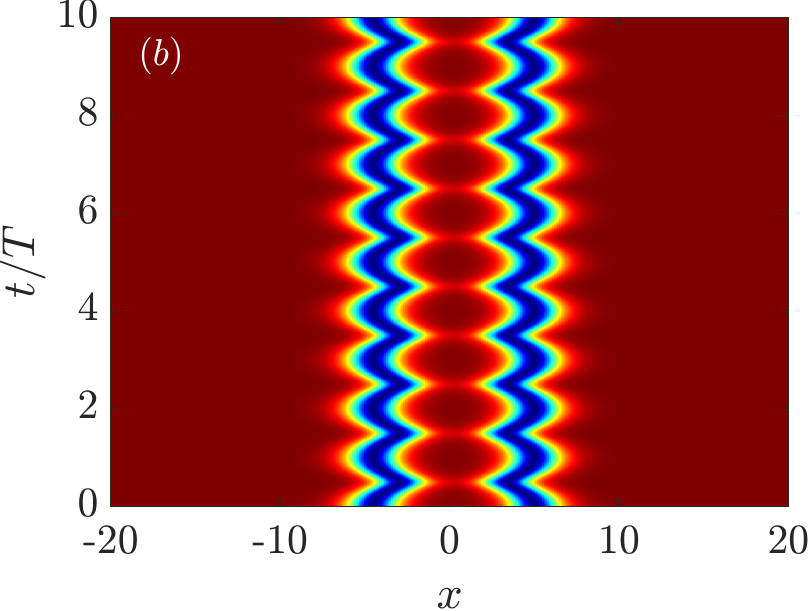} &
    \includegraphics[width=0.3\textwidth]{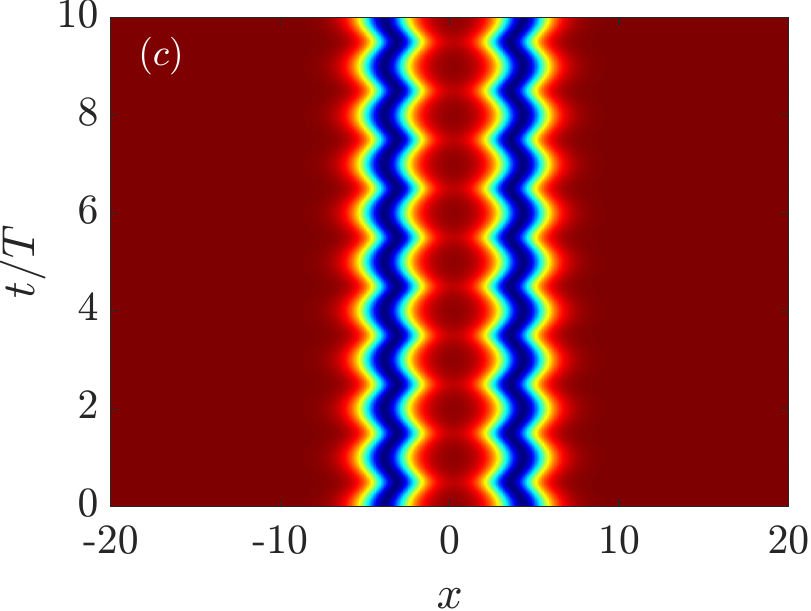} \\
    $\Omega=0.026$ & $\Omega=0.027$ & $\Omega=0.0271$ \\
    \includegraphics[width=0.3\textwidth]{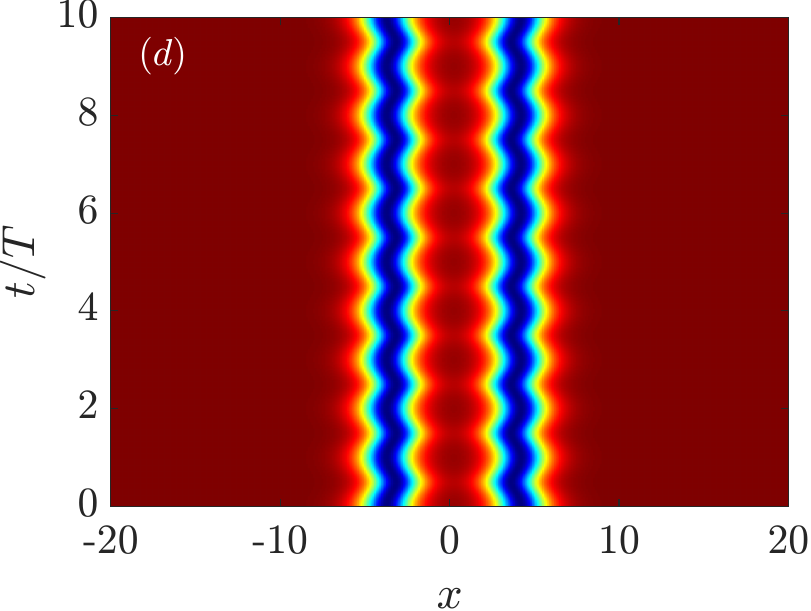} &
    \includegraphics[width=0.3\textwidth]{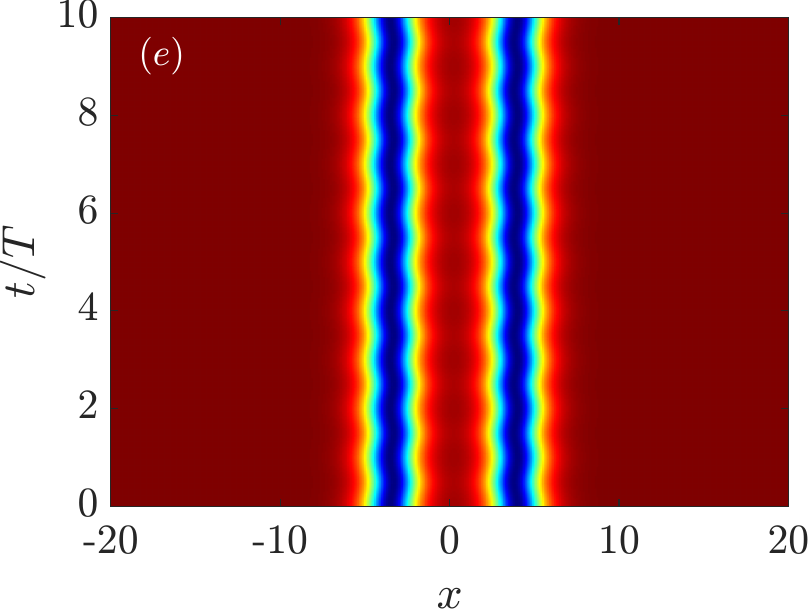} &
    \includegraphics[width=0.3\textwidth]{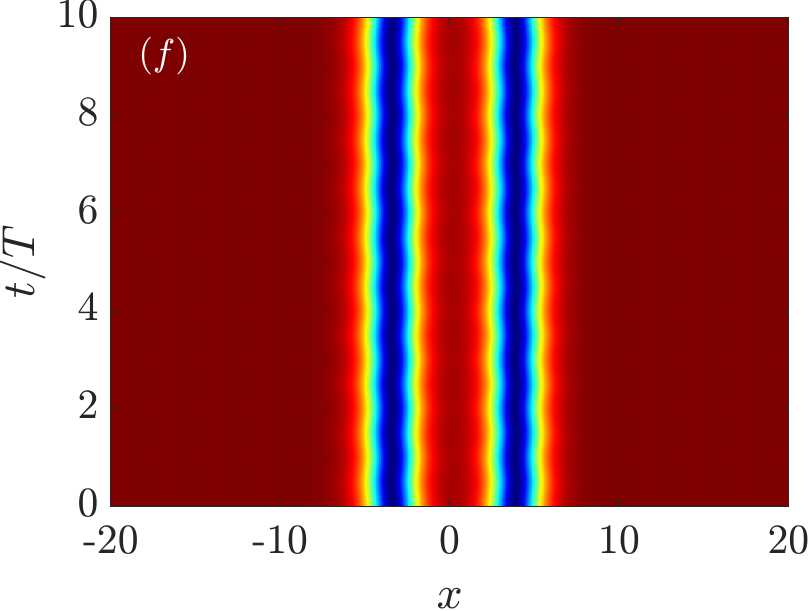} \\
\end{tabular}
    \caption{Evolution of the density $|\psi(x,t)|^2$ for breathers with $\alpha=2.01$. In every cases, ten oscillation periods $T$ are taken.}
    \label{fig:breatherexamp}
\end{figure}

The breathing character of the dynamical state stemming from an oscillatorily unstable bound state suggests the potential of time-periodic, spatially localized
breather states of  Eq.~(\ref{eq:Fourier}). We 
have succeeded in finding  such periodic orbits, e.g., for $\alpha=2.01$. To this effect, we had to increase the discretization parameter to $h=0.5$, and took the Fourier modes cutoff at $K=31$. {This high value of the discretization parameter is driven by the necessity of having a reasonable number of grid points for the Floquet analysis; otherwise, the integration times would be unreachable for our machines.} Figure~\ref{fig:breatherexamp} shows the evolution of these breathers for different values of $\Omega$.

{Breather solutions also present a hybridization with linear modes, similar to the ubiquitous nanoptera emerging in non-integrable continuous problems~\cite{KruskalSegur}. In principle, there is no lower bound for the breather frequency, although resonances with linear modes can make the continuation difficult when $\Omega\rightarrow0$. Interestingly, when $\Omega$ approaches $\Omega_c=0.027135687$, all the harmonics in Eq. (\ref{eq:Fourier}) tend to zero except for $k=-1$. Then, for $\Omega=\Omega_c$, the breather turns into a soliton pair with $\omega\rightarrow\omega-\Omega$ in a bifurcation reminiscent of a Hopf,
as a periodic orbit emerges from a stationary state. {In Fig.~\ref{fig:breatherexamp}, 
the space-time dynamics of the breather solutions are observed for
different frequencies, showcasing
how the breather transforms into a soliton molecule when $\Omega$ approaches $\Omega_c$.}

Finally, we have performed a Floquet analysis for such breathers finding again instability bubbles (see Fig.~\ref{fig:Floquet}(a)) although, contrary to bound states, these instabilities are of exponential nature, as shown in the example for the unstable breather depicted in the right panel of the same figure.

\begin{figure}
\centering
\begin{tabular}{cc}
    \includegraphics[width=0.45\textwidth]{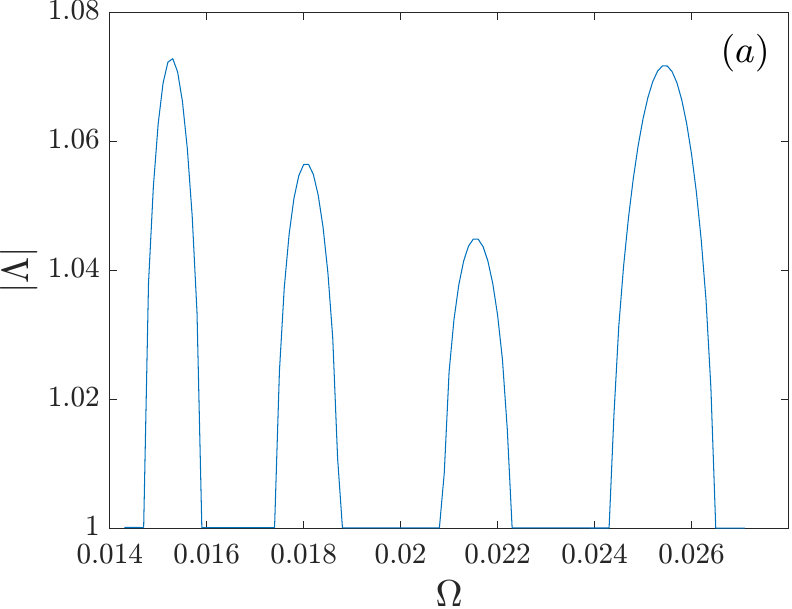} &
    \includegraphics[width=0.45\textwidth]{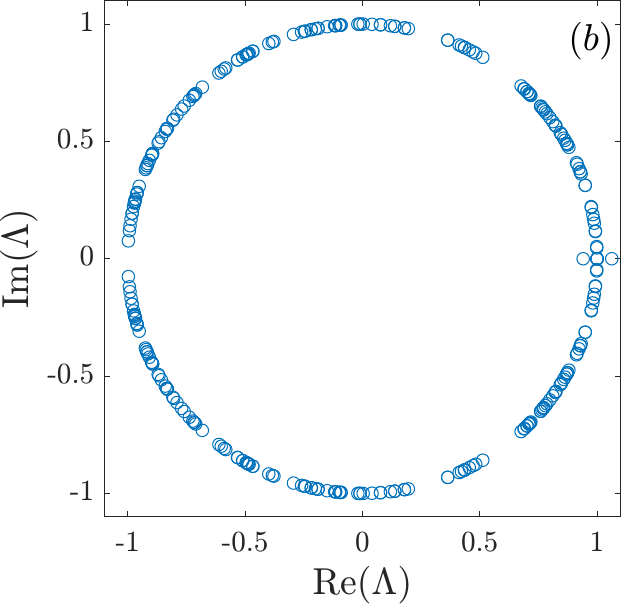} \\
\end{tabular}
    \caption{(a) Dependence of modulus of the Floquet multipliers with respect to $\Omega$ for breathers with $\alpha=2.01$. (b) Floquet multipliers spectrum for a breather with $\alpha=2.01$ and $\Omega=0.015$.}
    \label{fig:Floquet}
\end{figure}

In order to monitor the dynamics stemming from the instability of these breathers, we have perturbed the breather in Fig.~\ref{fig:Floquet}(b) following the direction of the unstable eigenvector. In that case, the dynamics is shown in Fig.~\ref{fig:breathersimul}, where one can observe that the effect of the instability is to induce  mobility to the breather.
The figure also shows the evolution of the projection onto the unstable eigenvector (\ref{eq:projection}); {in the case of breathers, $\lambda'=\Omega\log(\Lambda)/(2\pi\omega_0)$, with $\omega_0^2$ being equal to the wavefunction background, which is shifted by $-\Omega$; that is, $\omega_0^2=\Omega+|\omega|$.}

\begin{figure}
\centering
    \begin{tabular}{cc}
    \includegraphics[width=0.45\textwidth]{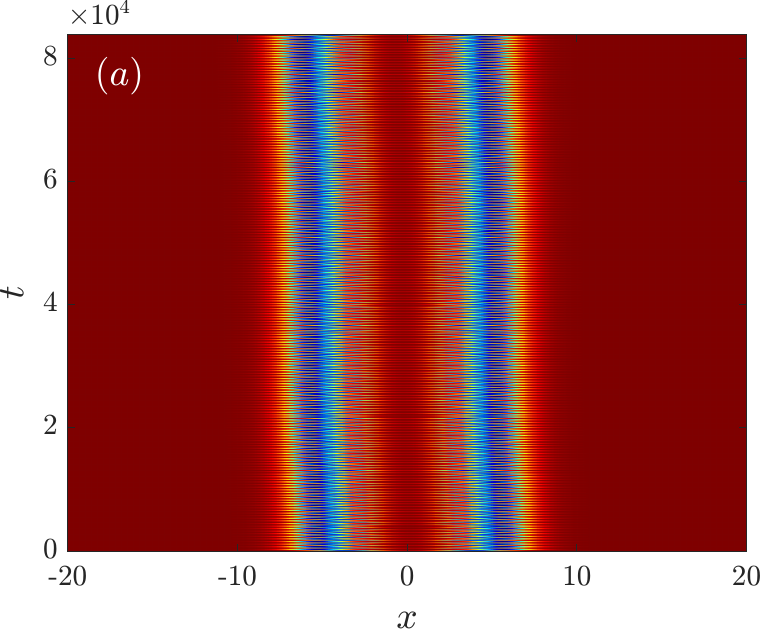} &
    \includegraphics[width=0.45\textwidth]{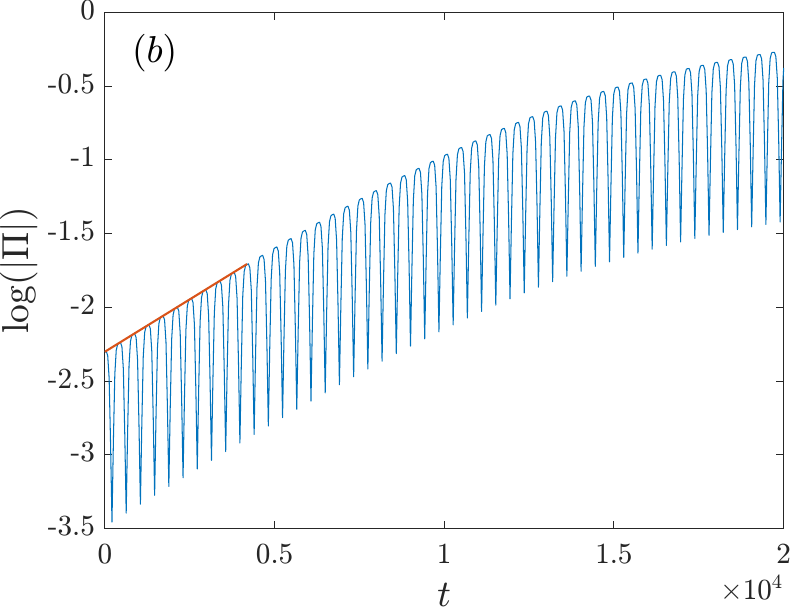} \\    
\end{tabular}
    \caption{Evolution of the density $|\psi(x,t)|^2$ of the breather with $\alpha=2.01$ and $\Omega=0.015$. (a) An spatio-temporal plot of $|\psi(x,t)|^2$. (b) The projection onto the unstable eigenvector as defined in Eq.~(\ref{eq:projection}). In this case, the unstable Floquet multiplier was $\Lambda=1.0628$ leading to a growth rate $\lambda'=1.44\times10^{-4}$; the value of the growth rate found from fitting the projection was $1.42\times10^{-4}$.}
    \label{fig:breathersimul}
\end{figure}

\section{Some Analytical Considerations}

In what follows we will provide two distinct attempts to understand
features of the states that our statics and dynamics have provided in
section III above. The first one will be a simpler ``one degree-of-freedom''
dynamical attempt to characterize the evolution of the inter-soliton
distance. The second one will be a more elaborate (in that it involves
three degrees of freedom) attempt to characterize the stationary states
involving the two kinks. While we do not seek to generalize
the latter to a dynamical form, it is a potential (more technically
involved) direction for future study.

\subsection{An ODE approach to the bound state dynamics}

The dynamics of the unstable bound states will be sought to be characterized via
a single ODE reduction. To this aim, we follow an approach similar to the one followed in \cite{Bob} for the fractional $\phi^4$ model, where a ``collective coordinate'' corresponding to the position of the density minimum, that will be denoted by $X$, behaves as a classical particle within a potential landscape.

To obtain the effective equation in the present NLS case, we consider non-stationary bound states with offset $\delta$ as minimizers of Eq. (\ref{eq:stat}) under the constraint $\phi(\delta)=0$ and taking Eq.~(\ref{eq:twosolitons}) as an initial seed. 
True stationary states exist {\it only} for the ``quanta'' of distances
discussed in the previous section. However, this optimization
procedure, occasionally referred to as ``vacuuming'', 
has been shown to produce suitable initial conditions for evolution dynamics,
especially so under long-range interactions~\cite{christov}.
Indeed, the resulting minimizer is used as initial condition for Eq. (\ref{eq:dyn}) and the evolution of the position of the density minimum $X(t)$ (with $X(t_0)=\delta$) is monitored. We restrict considerations to $X(t)>0$ as the solution is spatially even. To perform the calculations in this section, we have taken the domain $[-L,L]$, with $L=80$, and a discretization parameter $h=0.2${, although taking smaller values as $h=0.1$ gives similar dynamics}. The L\'evy index has been fixed to $\alpha=3.8$.

In the $\phi^4$ model considered previously,
the bound state either oscillates or is set dynamically into motion.  In the 
presently considered NLS problem, the dynamics is tantamount to an anti-damped particle in a potential landscape. This is caused by the existence of oscillatory instabilities which were not present in the 
real-valued $\phi^4$ model.
Indeed, this is reminiscent of what was examined in the work of~\cite{PelinovskyKevrekidis2008} where similar oscillatory instabilities
were obtained, and indeed a dynamical equation reminiscent of the one below
was suggested in the form of Eq.~(1.6) therein.

Here we posit that $X(t)$ follows the reduced ODE
\begin{equation}\label{eq:ODE}
    \ddot{X}-\gamma(X)\dot{X}-a(X)=0.
\end{equation}

{If we fix $\gamma(X)$ and $a(X)$ to the values for at a short time $t=t_0$, i.e., to $\gamma(\delta)$ and $a(\delta)$, we can integrate Eq.~(\ref{eq:ODE}) for (very) short intervals of time to obtain:
\begin{equation}
    X(t)=X(t_0)+\frac{\dot{X}(t_0)}{\gamma(\delta)}\left[\exp(\gamma(\delta) (t-t_0))-1\right]+ \frac{a(\delta)}{\gamma^2(\delta)}\left[\exp\left(\gamma(\delta)(t-t_0)\right)-\gamma(\delta)(t-t_0)-1\right].
\end{equation}
Then $\gamma(\delta)$ and $a(\delta)$ can be found by fitting $X(t)$ to the solution of (\ref{eq:ODE}). The interval in which the fit has been performed has been
empirically chosen to be from $t_0=0.3$ to $t_f=0.4$. The reason why fitting did not start from $t=0$ is because the existence of a (short) transient  ---due to the PDE discretization--- before the solution of Eq.~(\ref{eq:ODE}) can be applied.} 

Figure~\ref{fig:potential} shows the profiles of $a(\delta)$ and $\gamma(\delta)$, as they are obtained by repeating this fitting process (by means of a Levenberg-Marquardt optimization algorithm) for different distances $\delta$. At the moment, we do not have an interpretation for the functional form of the anti-damping, yet this is an interesting question for further investigation. {It may be of interest, accordingly, to explore avenues of 
perturbation theoretic formulations from the Laplacian limit, as has been done, e.g.,
in the context of potential perturbations of dark solitons in~\cite{PelinovskyKevrekidis2008}.} The space-dependent damping term suggests that Eq.~(\ref{eq:ODE}) represents a particle in an inhomogeneous viscous medium. The qualitative
interpretation of such an observation remains an interesting topic for further
exploration.

\begin{figure}
\centering
\begin{tabular}{cc}
    \includegraphics[width=0.45\textwidth]{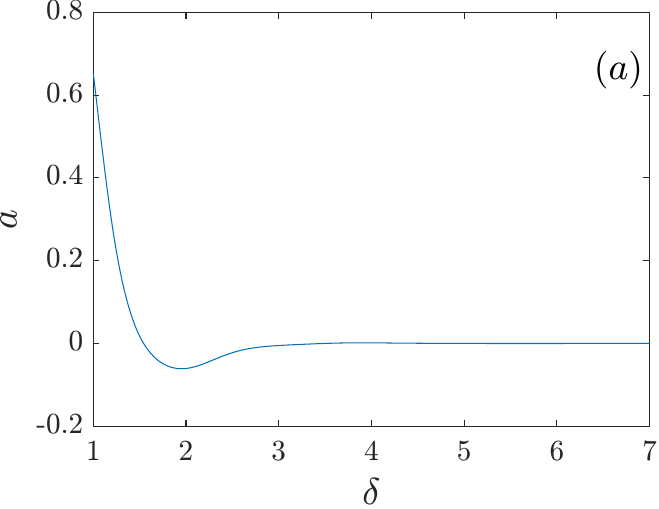} &
    \includegraphics[width=0.45\textwidth]{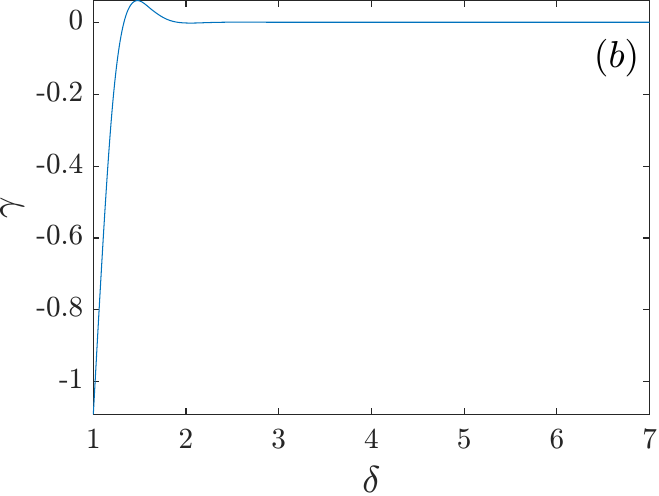} \\
\end{tabular}
    \caption{Acceleration (a) and damping (b) landscape for bound states with $\alpha=3.8$.}
    \label{fig:potential}
\end{figure}
Once $a(\delta)$ and $\gamma(\delta)$ are known, one can trace the phase portrait by taking some selected initial conditions and compare it with the phase portrait obtained by simulations of minimizers with the same initial $\delta$. This has been explored in Fig.~\ref{fig:phase_portrait} and illustrates that our effective ODE approach
provides a reasonable reduced phase portrait description of the two-soliton
effective dynamics. Clearly, while the relevant qualitative characterization matches
well that of the full fractional NLS model, obtaining a quantitative matching
(potentially so from first principles) is a particularly relevant objective for future
studies. Further analysis suggests that the main discrepancies between the two phase portraits appear when considering initial values of $X$ which are in the first potential minimum and close to the first maximum; in other words, the ODE approach is unable to provide a good approximation for small $\delta$, i.e., when the solitons comprising the bound state are too close and accordingly the repulsive force between them is strong. This is intuitively reasonable to expect as in this
limit the individual character of the solitary waves is less clear/pronounced.

\begin{figure}
\centering
\begin{tabular}{cc}
    \includegraphics[width=0.45\textwidth]{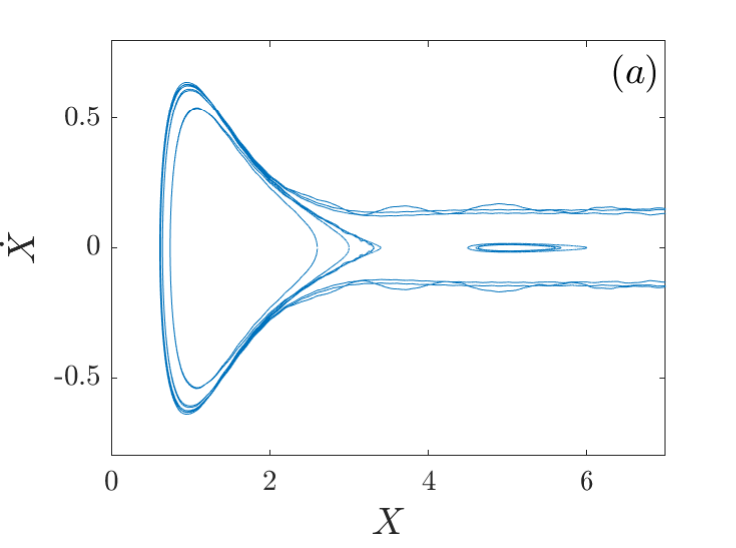} &
    \includegraphics[width=0.45\textwidth]{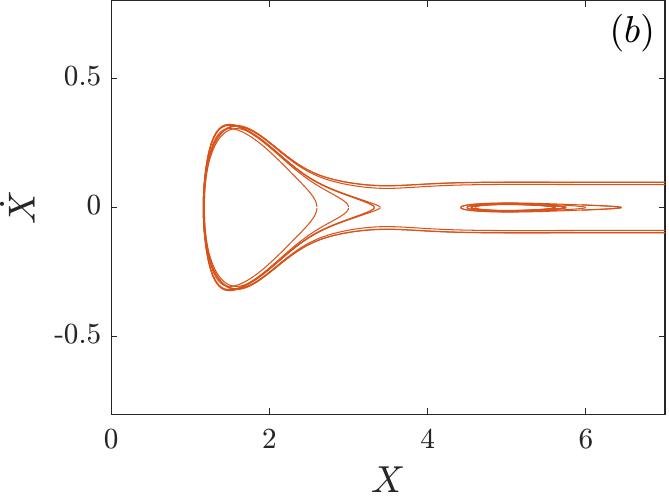} \\
\end{tabular}
    \caption{Phase portrait obtained (a) from the evolution of minimizers; (b) from the acceleration and damping landscapes of Fig.~\ref{fig:potential}.}
    \label{fig:phase_portrait}
\end{figure}

\subsection{Variational approach for stationary bound states}

We now turn to the analytical characterization of stationary states.
To that effect, we introduce the following ansatz in Eq.~(\ref{eq:stat}), fixing $\omega=1$ therein:
\begin{equation}\label{eq:ansatz}
    \phi(x)=A\mathrm{e}^{-ax^2}(1-bx^2)-1.
\end{equation}
The idea here is that we use an additional degree of freedom to obtain a 
more accurate description of the relevant waveform.
The advantage of this choice is that the Riesz fractional derivative of this function can be analytically calculated, giving

\begin{equation}
    \partial^\alpha_x\phi(x)=A a^{\frac{\alpha}{4}-1} \frac{\Gamma\left(\alpha/2\right)}{\Gamma\left(\alpha/4\right)} \left[(b-2 a) \, _1F_1\left(\frac{\alpha +2}{4};\frac{1}{2};-a x^2\right)-b(\alpha/2 +1) \, _1F_1\left(\frac{\alpha +6}{4};\frac{1}{2};-a x^2\right)\right]
\end{equation}
with $_1F_1\left(\cdot;\cdot;\cdot\right)$ being the confluent hypergeometric function (or Kummer function). Although it can expressed in terms of the parabolic-cylinder function, the resulting expression is less compact than the present one.
At the same time, the waveform also bears a disadvantage that we mention
for completeness, namely the relevant waveform (representing the numerical fractional NLS 
solution) for general $\alpha$ does {\it not}
decay exponentially; instead it decays as a power law. So the expression
of Eq.~(\ref{eq:ansatz}) is guaranteed to fail at capturing the far 
two-solitonic tails. A potential improvement that captures the correct
power-law structure, while remaining analytically tractable
is certainly desirable in future studies.

We define the same variational quantity as in Eq.~(4.1) of \cite{AtanasPanos}
\begin{equation}
\label{eq:variationalHu}
I[u(x)]=\frac{1}{2}\int_{-\infty}^{\infty} \left(\partial^{\alpha/2}_x u(x)\right)^2\mathrm{d}x+\frac{1}{4}\int_{-\infty}^{\infty}(1-u^2(x))^2\mathrm{d}x.
\end{equation}

When introducing the ansatz in the equation above, we get
\begin{equation}
\label{eq:variationalH}
I(A,a,b)=\frac{A^2\sqrt{\pi}}{27\times2^{16}a^{9/2}}\left[6\sqrt{2}a^2B_\alpha(a)\eta_{\alpha,1}(a,b)+\sum_{j=0}^4\Theta_j(A)a^{4-j}b^j\right].
\end{equation}

Upon minimization on variables $A$, $a$, $b$, the following set of algebraic equations is attained
\begin{equation}
\label{eq:variationaleq}
\begin{split}
B_\alpha(a)\eta_{\alpha,1}(a,b) &=\sum_{j=0}^4\zeta_j(A)a^{4-j}b^j, \\
B_\alpha(a)\eta_{\alpha,2}(a,b) &=\sum_{j=0}^4\xi_j(A)a^{4-j}b^j, \\
B_\alpha(a)\eta_{\alpha,3}(a,b) &=\sum_{j=0}^3\theta_j(A)a^{3-j}b^j.
\end{split}
\end{equation}

The functions appearing in Eqs. (\ref{eq:variationalH}) and (\ref{eq:variationaleq}) are given by the following relations:
\begin{equation*}
\begin{split}
B_\alpha(a)&=9\times2^{10}a^2\left(\frac{a}{2}\right)^{\alpha/2}\frac{\Gamma(\alpha)}{\Gamma(\alpha/2)}, \\
\eta_{\alpha,1}(a,b)&=16a^2+8(\alpha-1)ab+(\alpha^2+3)b^2, \\
\eta_{\alpha,2}(a,b)&=16(\alpha-1)a^2+8(\alpha^2-4\alpha+3)ab+(\alpha^3-5\alpha^2+3\alpha-15)b^2, \\
\eta_{\alpha,3}(a,b)&=4(\alpha-1)a+(\alpha^2+3)b; \\
\end{split}
\end{equation*}

\begin{align*}
\Theta_0&=-24\theta_1, & \Theta_1&=-36\theta_1, & \Theta_2&=6\theta_2, & \Theta_3&=-4\theta_3, & \Theta_4&=-63\theta_4, \\
\xi_0&=-4\theta_1, & \xi_1&=-6\theta_1, & \xi_2&=-5\theta_2, & \xi_3&=\frac{14}{3}\theta_3, & \xi_4&=\frac{9}{2}\theta_4; \\
\end{align*}

\begin{align*}
\zeta_0&=-2\zeta_1, & \zeta_1&=9\times2^{11}(\sqrt{2}A^2-2\sqrt{6}A+4), \\ \zeta_2&=-1152(9\sqrt{2}A^2+-16\sqrt{6}+24), &
\zeta_3&=160\sqrt{2}A(27A-32\sqrt{3}), \\ \zeta_4&=-945\sqrt{2}A^2; \\
\end{align*}

\begin{align*}
\theta_1&=1536(3\sqrt{2}A^2-8\sqrt{6}A+24), & \theta_2&=192(27\sqrt{2}A^2-64\sqrt{6}A+144), \\
\theta_3&=40\sqrt{2}A(81A-128\sqrt{3}), & \theta_4&=\zeta_4. \\
\end{align*}

We can find a solution for the triplet $(A,a,b)$ that is very close to the lower branch of solutions in Fig.~\ref{fig:bifbound} for large values of $\alpha$. This solution can be continued to lower $\alpha$ until the value $\alpha=2.52$ is reached. Although we cannot reach the limit of $\alpha=2$ for numerical bound states, the values of the amplitude ($\phi(0)$) and offset ($\delta$) predicted by the variational approach are close to those of the numerical solitons for $\alpha\gtrsim3$, as can be seen in Fig.~\ref{fig:variational1}. A comparison between the profile predicted by the variational approach and the numerical one is depicted in Fig.~\ref{fig:variational2}, where an excellent agreement is found for $\alpha=3.8$ (right panels (b) and (d)) 
and a quite reasonable one
is obtained even all the way down to $\alpha=2.6$ ((a) and (c)). 
While the ansatz is inadequate below $\alpha=2.5$, it provides a reasonable
semi-analytically tractable approximation in the interval of $[2.5,4]$
that can be used as a basis for further analysis of the relevant solitonic
molecules (i.e., bound states) and potentially of their dynamics.

As pointed out above, the variational approach can be improved by using an ansatz with a power-law decaying tail. To this aim, we tried to use hypergeometric functions having the same decay as the numerical bound state, but it gave rise to a variational equation set involving higher transcendental functions which did not shed any light to the problem. Moreover, the variational approach presented in this section fails when approaching the Laplacian limit, which suggests that the main problem in the ansatz lies in the core of the bound state rather than its tails. Nevertheless,  seeking  a suitable ansatz that improves the one presented herein and allowing one to find a suitable ODE reduction as well, is a crucial
challenge for the further consideration of variational approximations in this
field that is particularly worthwhile to consider in future studies.

\begin{figure}
\begin{center}
\begin{tabular}{cc}
    \includegraphics[width=0.45\textwidth]{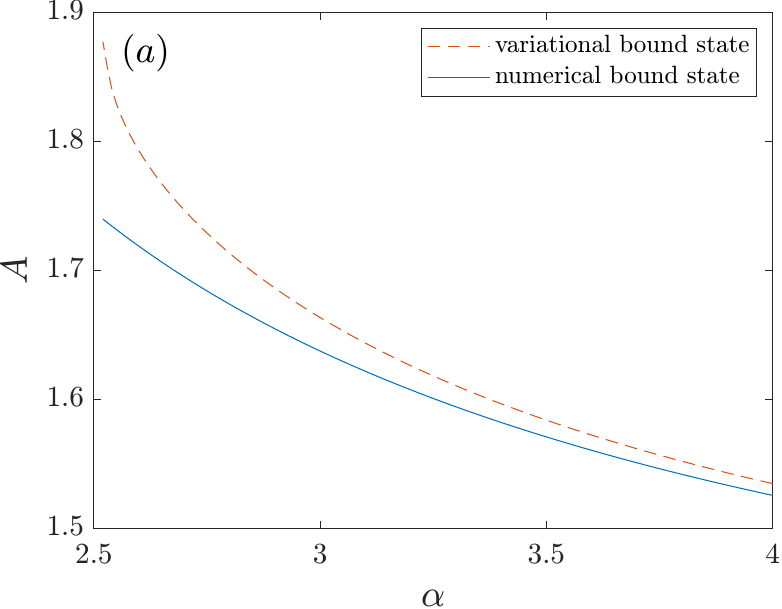} &
    \includegraphics[width=0.45\textwidth]{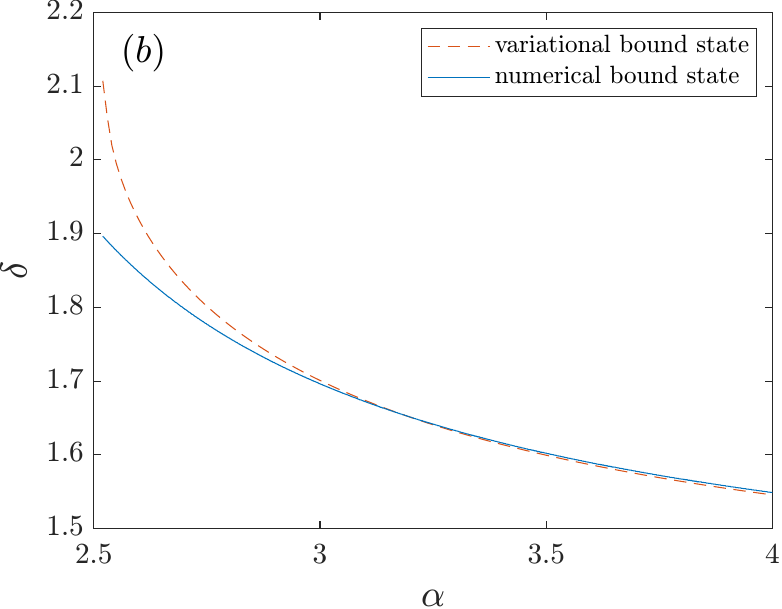}
\end{tabular}
\end{center}
\caption{Dependence of the amplitudes (a) and offsets (b) with respect to $\alpha$ for the numerical (blue solid line) and variational (red dashed line) bound states of the lowest branch in Fig.~\ref{fig:bifbound}.}
\label{fig:variational1}
\end{figure}

\begin{figure}
\begin{center}
\begin{tabular}{cc}
    \includegraphics[width=0.45\textwidth]{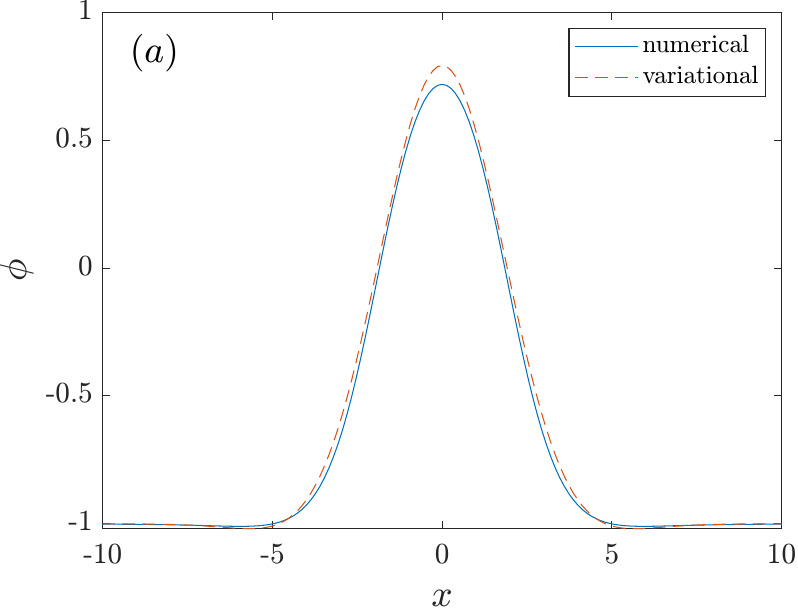} &
    \includegraphics[width=0.45\textwidth]{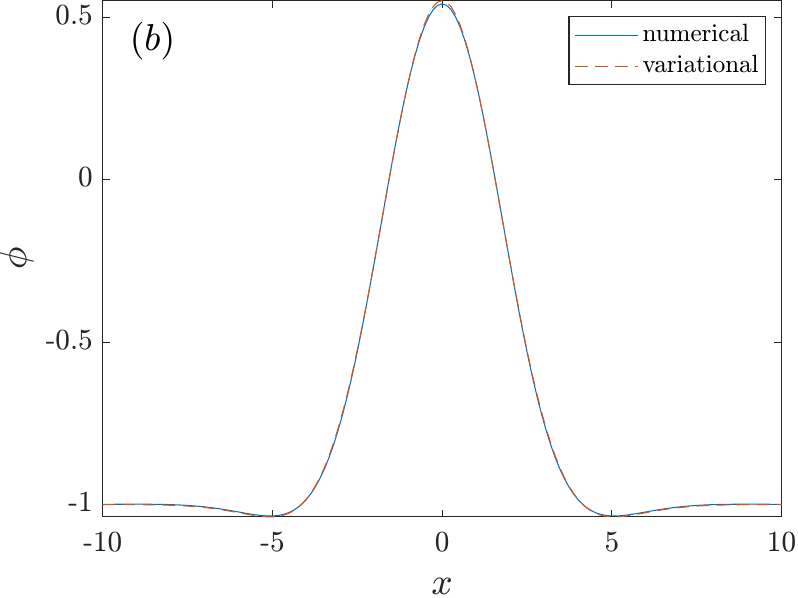} \\
    \includegraphics[width=0.45\textwidth]{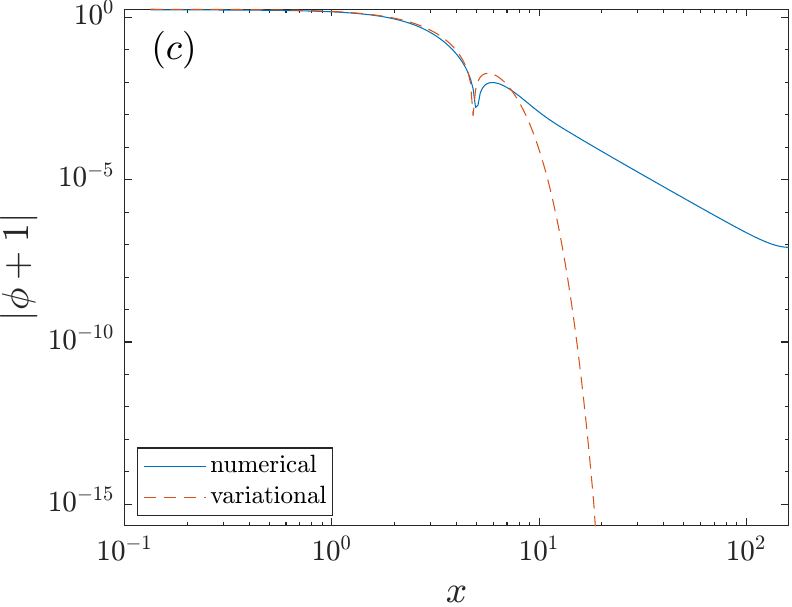} &
    \includegraphics[width=0.45\textwidth]{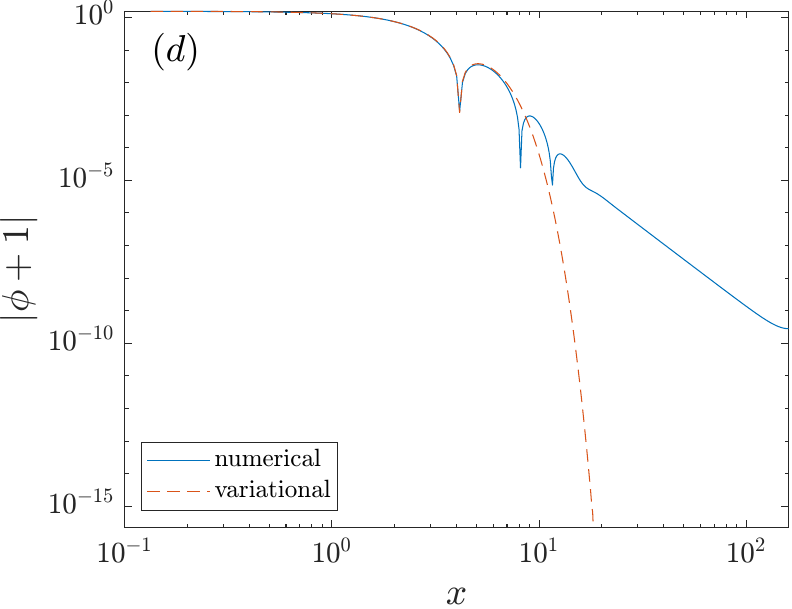}    
\end{tabular}
\end{center}
\caption{Profiles of the numerical (blue solid line) and variational (red dashed line) bound states of the lowest branch in Fig.~\ref{fig:bifbound} for $\alpha=2.6$ (a) and $\alpha=3.8$ (b). {The bottom panels (c) and (d) are the logarithmic plot counterparts of (a) and (b), respectively, showcasing how the core of the structure is qualitatively
(and indeed semi-quantitatively) captured, yet its tail is missed by the
ansatz.}}
\label{fig:variational2}
\end{figure}

\section{Conclusions and Future Challenges}

In the present work we sought to explore the dynamics of dark solitons in the
fractional NLS model. We are motivated by the recent experimental realization
in optical systems of arbitrary fractional dispersion operators~\cite{HoangNatCommun2025FractionalDerivative} for focusing nonlinear
settings, which suggests that defocusing corresponding implementations may
also be imminent. We discussed the existence and stability of a single
dark soliton under a Riesz-derivative induced fractional dispersion operator
parametrized by the L\'evy index $\alpha$ and then turned to the main
focus of the present work, namely the realm of solitonic pairs. 
We explained the similarities (as concerns the stationary state existence)
and significant differences (as concerns their stability and the potential
herein for oscillatory instabilities). The dynamics of the oscillatory
and exponential instabilities was also explored in the system dynamics
and, in turn, motivated the exploration of breathing states which were
computed using periodic orbit solvers, and their respective stability was
also examined via Floquet theory. Lastly, we sought to complement our
numerical findings using some semi-analytical tools. In particular, we 
provided a (data-informed) reduced order model description of the
soliton separation dynamics, which effectively describes the out-of-phase
motion of the solitonic molecules identified herein. We also considered
an analytically tractable ansatz, bearing three parameters which we 
could optimize to obtain a good fit to the steady states of the system
for a fairly wide interval of values of the L\'evy index $\alpha$.

Naturally, our studies pave the way for numerous future studies
some of which we have already highlighted herein. 
For instance, analytical approximations that may be tractable for
variational analysis, yet which respect the power-law decay of the
general L\'evy index case would be worthwhile to consider. 
Extending such considerations beyond static variants to fully
dynamic ones that can, in turn, be compared to the full fractional
NLS PDE findings, as well as to the data-driven
approximations thereof considered herein would be particularly
worthwhile. Indeed, in addition to being relevant for comparison
with dynamics, such approximations might be relevant for comparison
with the stability results presented herein, in analogy to what
was done in the work of~\cite{PelinovskyKevrekidis2008} (cf., 
e.g., the comparison of Eqs. (1.6) and (4.15) therein).
Furthermore, going beyond one-component, one-dimensional considerations
to either multi-component settings (involving, e.g., dark-bright 
solitons~\cite{buschanglin}), as well as to higher dimensional ones
involving vortical structures~\cite{siambook} and their characterization
would constitute fruitful directions for further exploration.
Such directions are currently in progress and will be reported in 
future publications.

{ {\bf Acknowledgements:} A.P.M. acknowledges support from the VII PPIT-US of the University of Seville under grant SOL2024-29847. This author also acknowledges the continued support from the research group FQM-201 from \textit{Junta de Andalucía}, Spain. J.C.-M. acknowledges support from grants PID2022-143120OB-I00 and CEX2024-001517-M, both funded by MICIU/AEI/10.13039/501100011033 and ERDF/EU.
This work was supported in part by the U.S. National Science Foundation under the award PHY-2408988
(P.G.K.). This research was partly conducted while P.G.K. was  visiting the Okinawa Institute of Science and
Technology (OIST) through the Theoretical Sciences Visiting Program (TSVP), the University of
Sydney through the visitor program of the Sydney Mathematical Research Institute (SMRI) and the Department of Mechanical Engineering at Seoul National
University through a Fulbright Fellowship. Their support is gratefully acknowledged.
Finally, this work was also  supported by a grant from the Simons Foundation [SFI-MPS-SFM-00011048, P.G.K.].
The authors also warmly acknowledge numerous helpful discussions and iterations with Bob Decker on
topics related to the present work.} 

{\bf Competing Interests:} The authors have no relevant financial or non-financial interests to disclose.

{\bf Data Availability:} Relevant data are available upon reasonable request from the authors.

\bibliographystyle{apsrev4-2}
\bibliography{Bibfile3}

\end{document}